\documentclass[a4paper,fleqn,usenatbib]{mnras}
\usepackage{mathptmx}
\usepackage[T1]{fontenc}
\usepackage{ae,aecompl}

\usepackage{graphicx}	% Including figure files
\usepackage{amsmath}	% Advanced maths commands
\usepackage{amssymb}	% Extra maths symbols

\newcommand{\mh}{m_{\rm H}}

\newcommand{\nh}{n_{\rm H}}

\newcommand{\hm}{{\rm H}_2}

\newcommand{\hmp}{{\rm H}_{2}^{+}}

\newcommand{\htwoo}{{\rm H}_{2}{\rm O}}

\newcommand{\hthreeop}{{\rm H}_{3}{\rm O}^{+}}

\newcommand{\hcop}{{\rm HCO}^+}

\newcommand{\percc}{\rm \,cm^{-3}}

\newcommand{\gpersqcm}{\rm \,g\,cm^{-2}}

\newcommand{\ps}{{\rm s}^{-1}}
\newcommand{\ccps}{{\rm cm}^{3}\, {\rm s}^{-1}}
\newcommand{\cmps}{{\rm cm} \, {\rm s}^{-1}}

\def\micron{\hbox{$\mu$m}}
\newcommand{\bcen}{\begin{center}}
\newcommand{\ecen}{\end{center}}
\newcommand{\be}{\begin{equation}}
\newcommand{\ee}{\end{equation}}
\newcommand{\bdis}{\begin{displaymath}}
\newcommand{\edis}{\end{displaymath}}

\title[Deep-Down Ionization of Protoplanetary Disks]{Deep-Down Ionization of Protoplanetary Disks}

\author[Glassgold, Lizano \& Galli]
{A.~E.~Glassgold$^1$\thanks{aglassgold@berkeley.edu}, S.~Lizano$^2$, D.~Galli$^3$\\
$^1$Astronomy Department, University of California, Berkeley, CA 94720, USA\\
$^2$Instituto de Radioastronom\'\i a y Astrof\'\i sica,  UNAM, 
Apartado Postal 3-72,  58089 Morelia, Michoac\'an, M\'exico\\
$^3$INAF-Osservatorio Astrofisico di Arcetri, Largo E. Fermi 5, 50125 Firenze, Italy\\}

\date{Accepted 2017 August 15. Received 2017 August 15; in original form 2017 June 29}

\pubyear{2017}

\begin{document}
\label{firstpage}
\pagerange{\pageref{firstpage}--\pageref{lastpage}}
\maketitle

\begin{abstract}
The possible occurrence of dead zones in protoplanetary disks subject
to the magneto-rotational instability highlights the importance of
disk ionization. We present a closed-form theory for the deep-down
ionization by X-rays at depths below the disk surface dominated by
far-ultraviolet radiation. Simple analytic solutions are given for
the major ion classes, electrons, atomic ions, molecular ions and
negatively charged grains. In addition to the formation of molecular
ions by X-ray ionization of $\hm$ and their destruction by dissociative
recombination, several key processes that operate in this region
are included, e.g., charge exchange of molecular ions and neutral
atoms and destruction of ions by grains. Over much of the inner
disk, the vertical  decrease in ionization with depth into the disk
is described by simple power laws, which can easily be included in
more detailed modeling of magnetized disks. The  new ionization
theory is used to illustrate the non-ideal MHD effects of Ohmic,
Hall and Ambipolar diffusion for a magnetic model of a T Tauri star
disk using the appropriate Elsasser numbers.
\end{abstract}


\section{Introduction} 

Twenty-five years ago Balbus \& Hawley (1991) proposed a little
known instability in rotating magnetized disks, the magnetorotational
instability (MRI), that might explain how accretion and turbulence
occur in protoplanetary disks (PPDs). Gammie~(1996) stressed how
important the level of ionization is for the operation of the MRI.
Assuming that PPDs are permeated by interstellar cosmic rays, he
showed how too little ionization can lead to so-called dead zones
where the MRI does not occur, usually close to the mid-plane. It
was then argued that the cosmic rays could easily be blocked by the
strong winds of young stellar objects, and that X-rays could ionize
PPDs (Glassgold, Najita \& Igea~1997).

Our understanding of disk ionization is still incomplete for a
variety of technical and physical reasons. Although considerable
progress has been made in both theoretical and observational studies
of the warm surface layers of PPDs irradiated by stellar far-ultraviolet
(FUV) radiation (reviewed by Najita \& \'Ad\'amkovics~2017), little
is known about regions closer to the mid-plane where no clear
diagnostics have been identified so far. The underlying physics of
both X-ray and cosmic-ray ionization of the inner disk is also
incomplete. The Monte-Carlo calculations of X-ray absorption and
scattering (e.g., Igea \& Glassgold~1999; Ercolano \& Glassgold~2013)
become inaccurate beyond a certain depth into the disk, e.g., for
vertical column densities $N_{\rm H}$ with log $N_{\rm H} > 25.25$.
In addition, there is the issue of the blocking of the cosmic rays
by a strong, asymmetric stellar wind. Cleeves et al.~(2013) first
approached this problem and obtained CR ionization rates 1000 smaller
than the interstellar value due to the modulation of the cosmic
rays by the T Tauri star wind.

Thermal-chemical models of the disks surface often involve hundreds
of species and thousands of reactions, and as such are unsuitable
for inclusion with 2-d and 3-d solutions of the non-ideal MHD
equations. This difficulty can be overcome by resorting to the
simplified treatment used by Oppenheimer \& Dalgarno (1974) for
ionization in dark clouds, and applied to PPDs by Ilgner \& Nelson
(2006a,b) and Bai \& Goodman~(2009). In this approach, the ionization
theory treats only a small set of generic species, e.g., molecular
ions like $\hcop$ and $\hthreeop$ (represented by $m^+$), heavy
atomic ions like Na$^+$ and Mg$^+$ (represented by $M^+$), and
charged grains ($g^-$). This formulation still allows for the proper
treatment of the destruction of molecular ions by including
charge-transfer to heavy atoms as well as dissociative recombination
with electrons, with the subsequent destruction of the heavy atomic
ions by recombination on charged grains.  We adopt this perspective,
following Ilgner \& Nelson and Bai \& Goodman.

We focus on the ionization of the inner part of a PPD below the
surface layer dominated by FUV ionization, and assume that the disk
has experienced significant grain growth. The goal is to track the
ionization down to the mid-plane, which we refer to as ``deep-down``
ionization. An important element of the theory is the approximate
power-law dependence of the X-ray ionization parameter, $\tilde{\zeta
} \equiv \zeta_X / \nh$ on vertical column density $N_{\rm H}$,
where $\zeta_X$ is the ionization rate and $\nh$ is the number
density of hydrogen nuclei. This leads to simple power laws for the
ion abundances, relations that can be easily accommodated in MHD
calculations of disks. We illustrate these results by considering
all three non-ideal MHD effects (Ohmic, Hall and Ambipolar) for a
weakly-magnetized disk model of Lizano et al.~(2016) for the T Tauri
star disk.

Although the theory presented in this paper provides a significant
simplification of PPD ionization, many unknowns remain. The exact
level of both X-ray and cosmic-ray ionization approaching close to
the mid-plane are uncertain, and the role of mixing needs to be
considered. In addition, the choice of important physical parameters
like the X-ray luminosity, the grain surface area, and the gaseous
heavy-element abundance can affect the character of the ionization.
The importance of disk ionization goes beyond considerations of the
MRI because in general it determines the coupling of the field to
the disk ions and the coupling of the ions to the neutral species,
essential issues for any theory of magnetized disks.

\section{Physical Processes and Solution}

We apply the Oppenheimer \& Dalgarno (1974) method to PPDs by first
writing balance equations for the number density of molecular and
atomic ions, $n(m^+)$ and $n(M^+)$,
\be
\zeta_X \nh = (k \, n_M + \beta n_e + k_d n_d)\,n(m^+), 
\label{eq:bal1}
\ee
\be
k_{\rm ce} n_Mn(m^+) = (\alpha \, n_e + k_d n_d) \, n(M^+).
\label{eq:bal2}
\ee
Eq.~(\ref{eq:bal1}) expresses the idea that the ionization is due
to X-ray ionization of $\hm$ and that the proton in $\hmp$ is rapidly
transferred to high-abundance molecules that better bind the proton,
especially CO and $\htwoo$.  The right side of the equation states
that the molecular ions can be lost by charge exchange with neutral
atoms, dissociative recombination with electrons, and recombination
with grains. In this equation, the rate coefficient, $k=k_{\rm ce}
+ k_{\rm pt}$, is the sum of charge-exchange and proton-transfer
rate coefficients for atomic ions reacting with molecular hydrogen;
$n_M$, $n_e$ and $n_d$ are the number density of neutral atoms,
electrons, and dust, respectively.  Eq.~(\ref{eq:bal2}) describes
how atomic ions are produced by charge exchange of molecular ions
with neutral atoms and how they can be destroyed by either recombination
with grains or electrons. The rate coefficients $\alpha$ and $\beta$
have their usual values (Spitzer 1978, Anicich 1993),
\be
\alpha = 2 \times 10^{-10}\, T^{-2/3}\, \ccps,  \,
\beta = 1.2 \times 10^{-7}\, T^{-3/4}\, \ccps.
\ee     
The information available on the reaction of molecular ions with
atoms is limited, and we estimate that the rate coefficient for
charge exchange and proton transfer are about the same and equal
to,
\be
k_{\rm ce} = k_{\rm pt} = 5 \times 10^{-10}\, \ccps.
\ee

The grains are treated according to the collisional-charging theory
of Draine \& Sutin~(1987, hereafter DS87). For a typical grain size
$a$ (e.g., the geometric mean of the minimum and maximum sizes in
the Mathis et al.~1977 $a^{-3.5}$ power law distribution\footnote{We
adopt $10\,\micron$ as representative grain size, but the formulae
allow for consideration of other sizes.}), DS87 introduce an
``effective temperature''
\be 
\tau = \frac{a\,k_{\rm B}T}{e^2} = 59.9 \, 
\left(\frac{a}{10~\micron}\right)\; \left(\frac{T}{100~{\rm  K}}\right), 
\label{eq:Teff}
\ee
and an ``effective atomic weight'' of the ions $\mu_i$ defined by
their Eq.~(4.8b).  Assuming an ``electron sticking coefficient''
$s_e=1$, $\mu_i$ is given by
\be
\mu_i \equiv  \left(\frac{n_e}{n_i}\right)^{2} \left( \frac{m_i}{m_{\rm H}}\right), 
%m_{\rm ion} =    \left(\frac{n_{\rm ion}}{n_e}\right)^{2} \mu_i m_H,
\ee
where the total ion density is 
\be
n_i = n(M^+) + n(m^+),
\ee
and the average ion mass $m_i$ is given by 
\be
\frac{n_i}{m_i^{1/2}} =   \frac{n(M^+)}{m_M^{1/2}}  +  \frac{n(m^+)}{m_m^{1/2} } .
\ee
Assuming an ion mass, $m_k = A_k m_{\rm H}$, with $k=M, m$,
%and molecular ions, $m_M = 24 m_H$ (Na) and $m_m = 20 m_H$ (HCO$^+$),
the effective atomic weight can be written as
\be
\mu_i = {n_e}^2  \left[  \frac{n(M^+)}{A_M^{1/2}} +    \frac{n(m^+)}{A_m^{1/2}}          \right]^{-2}
\ee
The grain charge depends only on $\tau$ and $\mu_i$.  While $\tau$
depends on grain size and temperature, $\mu_i$ is bound by two
limiting values: if dust plays no role in the charge balance, $n_i
\approx n_e$ and $\mu_i \approx \langle A_i\rangle$, where $\langle
A_i\rangle \approx 25$.  Instead, if the grains carry free charge,
as a consequence of balance between negatively and positively charged
grains, $n_i/ {n_e} \approx (\langle A_i \rangle m_{\rm H}/m_e)^{1/2}$
and $\mu_i \approx m_e/m_{\rm H}$ (see, e.g., Eq.~28 of Nakano et
al.~2002).  In the limit of the big grains/high temperatures of
interest here ($\tau \gg 1$) the grain charge is given by
\be
\langle Z_d \rangle \approx -\tau  \psi ,
\ee
where $\psi$ is the solution of the charge equation (Spitzer 1941)
%%% Do we need the reference to Spitzer. Isn’t this equation given by DS87?
\be
(1+\psi) e^{\psi} = \sqrt{ \frac{\mu_i m_{\rm H}}{m_e}},
\ee
and $m_e$ is the mass of the electron. For electron and heavy atoms
with $m_M = 25 \, \mh$, $\psi = 3.8$, but if grains dominate  the
negative charge, $\psi$ will be much smaller. In the example treated
below, $\psi < 1$ near the mid-plane.  In applying the DS87 theory
to the present problem, we assume that the grains and the temperature
are large enough  for the effective temperature in Eq.~(\ref{eq:Teff})
to satisfy $\tau \gg 1$.

The contribution of the grains to the total charge fraction ($g^-$) is then,
\begin{eqnarray}
\nonumber
x(g^-)  =    &  -x_d \langle Z_d \rangle  \quad \quad \quad \quad \quad \quad \quad \quad \quad \quad \quad  
\quad \quad \quad  \\
  = & 1.12  \times 10^{-16} \; 
\left(\frac{\rho_g / \rho_ d}{0.01}\right) \, 
\left(\frac{10\,\micron}{a}\right)^2 \; \left(\frac{T}{100\,{\rm K}} \right) \, \psi, 
\label{eq:grcharge}
\end{eqnarray}
where the dust abundance relative to the density of H nuclei is,
\be
x_d =  \frac{n_d}{n_{\rm H}} 
 = 1.863 \times 10^{-18}\; \left(\frac{\rho_g / \rho_ d}{0.01}\right) \, 
\left(\frac{a}{10\,\micron}\right)^{-3}.
\label{eq:xd}
\ee

DS87 also give approximate expressions for the recombination rate coefficient $k_d$ for 
the reaction of heavy ions, both atomic and molecular, with grains,
\be
k_d \approx  \pi a^2 \langle v_i\rangle  \left ( 1 + \psi \right),
\ee
where the mean thermal ion speed is $\langle v_i\rangle  = \sqrt{8
k_{\rm B} T / \pi \langle A_i\rangle m_{\rm H}}$.  This rate can
be written as
\be
k_d = 9.11 \times 10^{-2} \; \left(\frac{a}{10\,\micron}\right)^2 \, 
\left(\frac{T}{100\,{\rm K}}\right)^{1/2} (1 + \psi)~\ccps.
\ee
The grain destruction term in eqs. (\ref{eq:bal1}) and (\ref{eq:bal2}) is then,
\be
k_d\, x_d = 1.70 \times 10^{-19} D~\ccps,
\label{eq:grd}
\ee
with
\be
D = \left(\frac{\rho_g / \rho_ d}{0.01}\right) \, \left(\frac{10\,\micron}{a}\right) \,  
\left(\frac{T}{100\,{\rm K}}\right)^{1/2} . %T^{1/2} \,(1 + \psi). 
\ee

The basics of the ionization theory can now be completed with the
equation for charge conservation,
\be
n(m^+) + n(M^+) = n_e + n(g^-).
\ee
It is then straightforward to obtain formulae for abundances of 
$m^+$ and $M^+$,
\be
x(M^+) = \frac{k_{\rm ce}\,x_M[x_e + x(g^-)]}{k_{\rm ce}\,x_M + k_d\,x_d + \alpha \, x_e},
\label{eq:hion}
\ee 
\be
x(m^+) = \frac{(k_d\,x_d + \alpha x_e)[x_e + x(g^-)]}
{k_{\rm ce}\,x_M + k_d\,x_d + \alpha x_e},
\label{eq:molion}
\ee
and then obtain a cubic equation for the electron fraction, 
\begin{eqnarray}
%\hspace*{0.5in} 
\lefteqn{\alpha \beta x_e^3 
+ [\alpha(k\,x_M + k_d\,x_d)  + \beta x_d (k_d - \alpha \langle Z_d \rangle) ]x_e^2} \nonumber \\
&  - \{\tilde{\zeta}\alpha -  k\,x_M \, x_d (k_d - \alpha \langle Z_d \rangle)  
+ x_d^2\,[ (\alpha + \beta)\,\langle Z_d \rangle - k_d^2 \;] \}x_e \nonumber \\ 
& - \,[\tilde{\zeta}(k_{\rm ce}\,x_M + k_d\,x_d) + k_d\,x_d^2\, k\,x_M\,\langle Z_d\rangle
+ k_d^2\, x_d^3 \, \langle Z_d\rangle] = 0.
\label{eq:cubic}
\end{eqnarray}

In principle, Eq.~(\ref{eq:cubic}) can be adapted to treat high as
well as low ionization regions of PPDs. The ionization rate $\zeta_X$
would then have to include a broader range of external radiation
than just the energetic X-rays considered here. In particular, the
surface layers are strongly affected by FUV radiation, including
Lyman-$\alpha$. As shown in recent models of the inner surface
layers, e.g., \'Ad\'amkovics et al.~(2014, 2016), shielding by dust
and molecules play an essential role in determining the effects of
FUV irradiation. In this report, we focus on that part of the PPD
that lies below the FUV layer where the ionization is many orders
of magnitude less than near the surface. In this deep-down region
of a PPD, the X-ray ionization rate plays a key role in determining
the ionization. It enters the cubic equation Eq.~(\ref{eq:cubic})
through the ionization parameter, $ \tilde{\zeta}$. As discussed
in earlier work (Igea \& Glassgold~1999; Ercolano \& Glassgold~2013),
the penetration of the X-rays to large depths depends on their
scattering by electrons, both free and weakly bound. These studies
show that the X-ray ionization parameter in the inner PPD (out to
$R \approx $  15-20 AU) decreases rapidly with height and relatively
slowly with radius. For the MMSN, the decrease follows the $-8/3$
power of the vertical column density $N_{\rm H}$ (measured from the
top),
\be
\tilde{\zeta} = \zeta_X / \nh = 10^{-33}\,  (N_{\rm H} / \,10^{26} \, 
{\rm cm}^{-2})^{-8/3} \,  \ccps.
\label{eq:ionpar}
\ee
This behavior is also expected to apply to other density
distributions\footnote{The slow decrease of $\tilde{\zeta}$ with
$R$ may be understood from the fact that both $\zeta_X$ and $\nh$
decrease as the inverse square of the radius. The deep-down variation
of $\tilde{\zeta}$ with column density in Eq.~(\ref{eq:ionpar})
stems from the fact that $\zeta_X$ varies with $N_{\rm H}$ slightly
less strongly than $R^{-2}$ and $\nh$  varies with $N_{\rm H}$
slightly less strongly than $R^{-1}$.}. However, our understanding
of X-ray ionization is incomplete at the largest depths in PPDs
(approaching the mid-plane) because of the limitations in the Monte
Carlo scattering calculations. In practice this means that the X-ray
ionization rates are uncertain for vertical columns greater than
log $N_{\rm H} = 25.25$.

The ionization level in the regions of interest for this work are
small, $ < 10^{-9}$, and even smaller approaching the mid-plane.
Under these conditions, the quadratic and cubic terms in
Eq.~(\ref{eq:cubic}) can be ignored, as can the quadratic and cubic
terms in $x_d$ in the last two terms of the equation. In addition,
the terms involving the radiative recombination coefficient $\alpha$
can be dropped, leading to a simple expression for the electron
fraction deep down in the inner part of a PPD,
\be
x_e \approx
  \tilde \zeta\, \frac{k_{\rm ce} \, x_M + k_d\,x_d}
			    {k\,x_M\, k_d\,x_d}.
\label{eq:xelred}
\ee
The relative size of the two terms in the numerator depend on the
abundance of heavy atoms in the gas phase ($k_{\rm ce} \, x_M$) and
on the abundance and size of the grains ($x_d\,k_d$ from
Eq.~\ref{eq:grd}). To estimate the order of magnitude of these
terms,  Eq.~(\ref{eq:xelred}) can be rewritten using Eq.~(\ref{eq:grd})
with $T = 100$\,K and $\rho_g/\rho_d = 0.01$, as
\be
\frac{k_{\rm ce} \, x_M + k_d\,x_d}{k\,x_M \,k_d\,x_d}
%= \frac{0.5}{k_d\,x_d} + \frac{1}{k\,x_M}
\approx \left[ \frac{2.9 \times 10^{18}}{1+\psi} \, 
\left(\frac{a}{10\,\micron}\right) + \frac{10^{9}}{x_M} \right] \, {\rm cm^{-3} \, s}.
%\frac{1}{\ccps}.
%\, \left(\frac{a}{10 \micron}\right)
\label{eq:ratio}
\ee
The abundance of heavy atoms $x_M$ determines the size of the second
term of this equation. Other than several well studied lines of
sight for diffuse clouds (e.g. Savage \& Sembach 1996), little is
known about the abundance of gas phase heavy atoms in thicker clouds
and protoplanetary disks. In diffuse clouds, the volatile atoms Na
and K are depleted by  about 1 dex, Mg and Si by 1.5 dex, the more
refractory Fe by $\sim 230$, and essentially no depletion for S. A
study of abundances for five thicker clouds by Joseph et al.~(1986)
yielded a larger depletion for Si and a 1 dex depletion for S. More
recently, Anderson et al.~(2013) measured the $25\, \micron$ S\,I
fine-structure emission from three Class\,0 sources with outflows
and shocks. They obtain lower and upper limits for the depletion
of atomic S, which they interpreted as produced by shock-induced
vaporization of sulfur-rich grain mantles.  There is no new information
on the abundance of Na and K. We adopt a conservative approach to
this fragmentary observational situation by adopting the largest
depletion factors suggested for Na (10), Mg (40), Si (75), S (100),
K (12) and Fe (180). The average abundance of these heavy atoms is
then $x_M = 1.8 \times 10 ^{-6}$, close to the undepleted, solar
Na abundance. This number might well be smaller because it relies
heavily on the uncertain depletions of Mg and Si. Our best conservative
guess is $x_M = 10 ^{-7}$, or 0.1\% of the solar abundance. If we
use this value in Eq.~(\ref{eq:ratio}), we find that the second
term is between 1 and 2 dex smaller that the first, unless the
effective grain size (surface area) is very small, i.e, less than
$0.1 \micron$. Where the second term can be ignored, Eq.~(\ref{eq:xelred})
becomes,
\be
x_e \approx \tilde \zeta \, \frac{k_{\rm ce} / k}{k_d\,x_d}=0.5\frac{\tilde\zeta}{k_d\,x_d}, 
\label{eq:simplest}
\ee
directly expressing how the electron fraction is determined by the
ion production by X-rays and destruction by grains (see also Eq. (58) of Okuzumi \& Inutsuka 2015). This equation
should suffice for a preliminary discussion of deep-down ionization.

\section{Results for Ion Abundances}

\begin{figure}
\centering
\includegraphics[angle=0,width=\columnwidth]{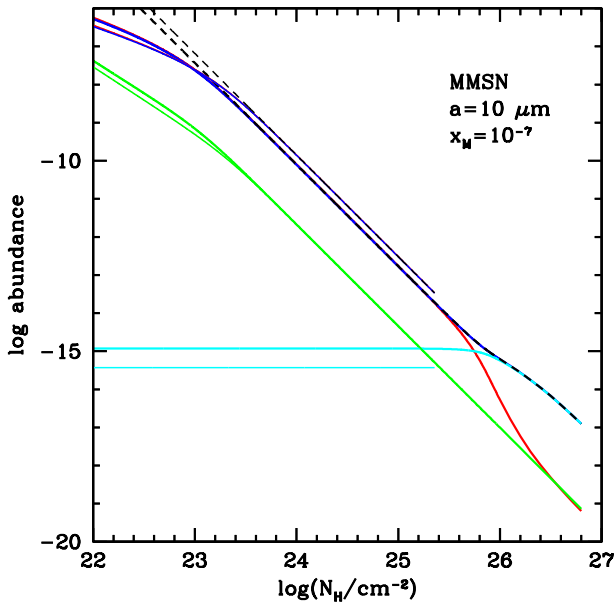} 
\caption{ Abundances as function of vertical column density for a MMSN disk at $R = 1$ AU and $10$ AU,
for grains with size $a = 10\, \micron$, and heavy atom abundance $x_M = 10^{-7}$.
The curves are for $x_e$ ({\it red}\/), $x(M^+)$ ({\it blue}),
$x(m^+)$ ({\it green}\/), and $x(g^-)$ ({\it cyan}\/). The {\it
dashed black curves} correspond to the simple analytic solution,
Eq.~(\ref{eq:simplest}). The thicker lines correspond to $R = 1$ AU; these lines 
reach the highest column density ($N_H = 7.2 \times 10^{26}
\, {\rm cm^{-2}}$).  The curves shift slightly
with $R$ in accord with the decline in the temperature of the
isothermal atmosphere  of the MMSN.}
\label{fig:10m_1AU_10AU}
\end{figure}

We illustrate the theory presented in the previous section by
applying it to the Minimum Mass Solar Nebula (MMSN,
Hayashi~1981)\footnote{The MMSN parameters at 1\,AU are: temperature,
280\,K; scale height, $5\times 10^{11}$\,cm; and mid-plane density,
$1.15 \times 10^{15} \, \percc$; the mass column of H nuclei is
then $1700\,\gpersqcm$; the mid-plane temperature and density vary
with radius as the $-0.5$ and $-2.75$ powers, respectively.}.  The
reference case is defined by a grain size of $a = 10\,\micron$ and
a heavy atom abundance $x_M= 10 ^{-7}$. {Figure~\ref{fig:10m_1AU_10AU}
shows the solution of the cubic equation Eq.~(\ref{eq:cubic}) for
a disk radius $R =1$\,AU  with abundances plotted against vertical
column density in the range log $N_{\rm H} = 22$--26. The main ions,
$e$, $M^+$, $m^+$ and $g^-$, all follow power laws to a good
approximation. The ionization is largely dominated by the relation
$x_e = x(M^+)$. The molecular ions follow a parallel line but down
by 1.5 dex. The negative grain abundance in this case is approximately
constant at $x(g^-) \approx 10^{-15}$. It plays no role until close
to the mid-plane of the disk where it becomes more abundant than
electrons. The simple analytic abundance formula, Eq.~(\ref{eq:simplest}),
is represented by a dashed line. It agrees closely with the solution
to the cubic except at small columns where the straight line fit
begins to fail, and also at the largest columns where negatively-charges
grains replace electrons as the major carrier of negative charge.
The relatively high electron abundance at small columns is due to
X-ray ionization producing high abundances of heavy atomic ions,
the basis for the present theory of ionization deep down. In a more
complete theory of disk ionization, $x_e$ will be even larger at
small columns due to FUV ionization. The electron fraction generally follows
the $-8/3$ power dependence of the ionization parameter $\tilde{\zeta}$
in Eq.~(\ref{eq:ionpar}).
Figure~\ref{fig:10m_1AU_10AU} also shows the ion abundances for
the disk radius $R = 10$\,AU. There is a
small decrease in ionization with increasing radius.  This is due
to the decline with radius of the temperature of the isothermal
atmospheres in the MMSN.  The numerator in the second factor of
Eq.~(\ref{eq:simplest}) is assumed to be fixed at 0.5, and the
denominator $k_d\,x_d$ does not change with grain size because grain
size has been fixed in this figure.  The term $k_d\,x_d$ does produce
the small variation with radius via its temperature dependence, according
to Eq.~(\ref{eq:grd}). The grain charge also decreases with decreasing
temperature, according to Eq.~(\ref{eq:grcharge}). }

The behavior of the ions can be understood by examining Eq.~(\ref{eq:hion})
and Eq.~(\ref{eq:molion}) for $M^+$ and $m^+$. The denominators of
both equations are dominated by the first term, $k_{\rm ce}\,x_M$.
Applying $x(g^-) \ll x_e$ in the numerator of Eq.~(\ref{eq:hion}),
then leads to $x(M^+) = x_e$, in agreement with the figure. Ignoring
small terms in the numerator of Eq.~(\ref{eq:molion}) then leads to,
\be
x(m^+) = \frac{k_d\,x_d}{k_{\rm ce}\, x_M}\,x_e,
\ee
so that $x(m^+)$ is both proportional to but much smaller than
$x_e$, as shown in the figure.

\begin{figure}
\begin{minipage}[c]{\columnwidth}
\includegraphics[angle=0,width=\columnwidth]{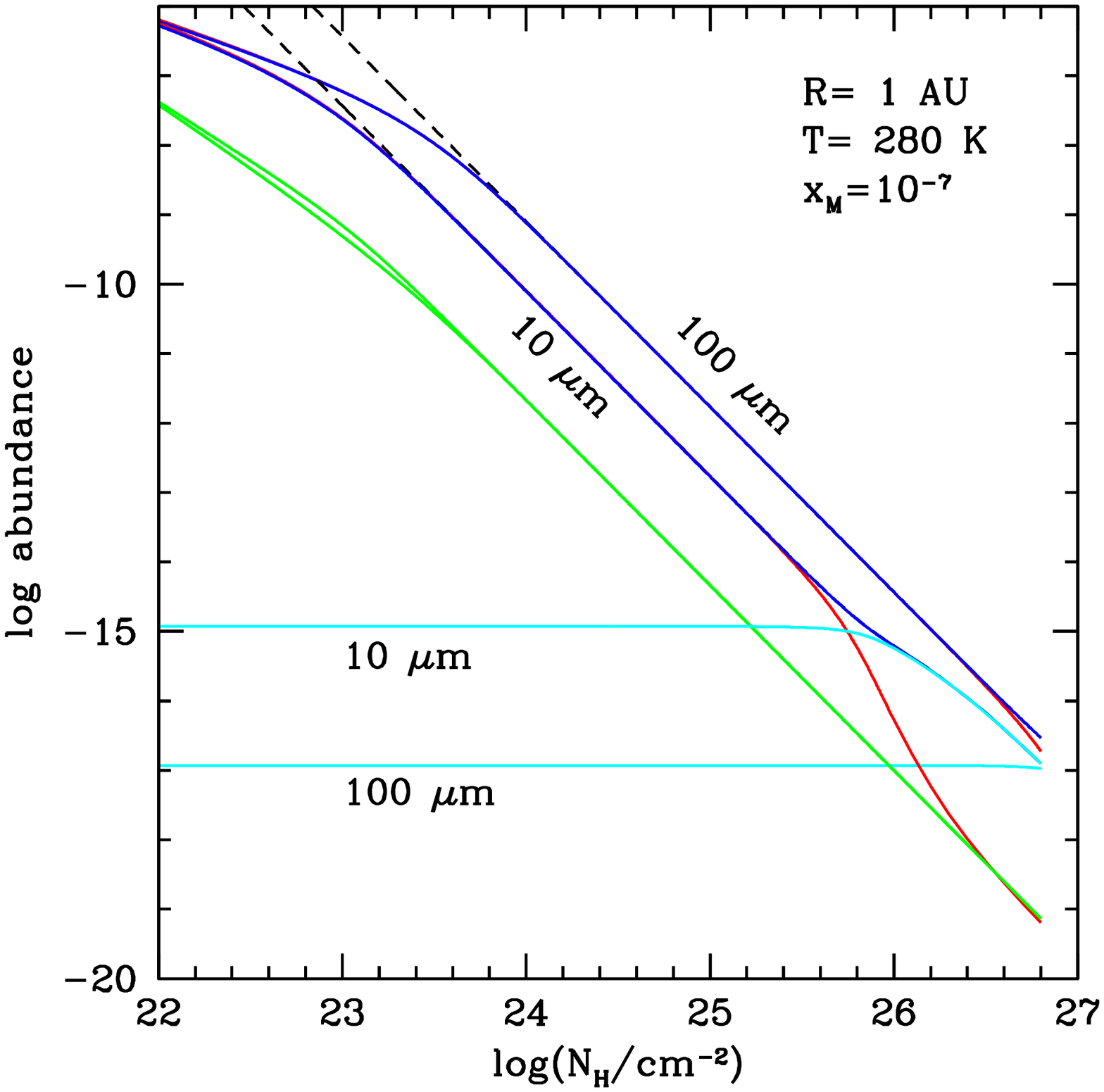}  %{summ_1AU_10_100mu.eps}
\end{minipage}
\hspace{0.5cm}
\begin{minipage}[c]{\columnwidth}
\includegraphics[angle=0,width=\columnwidth]{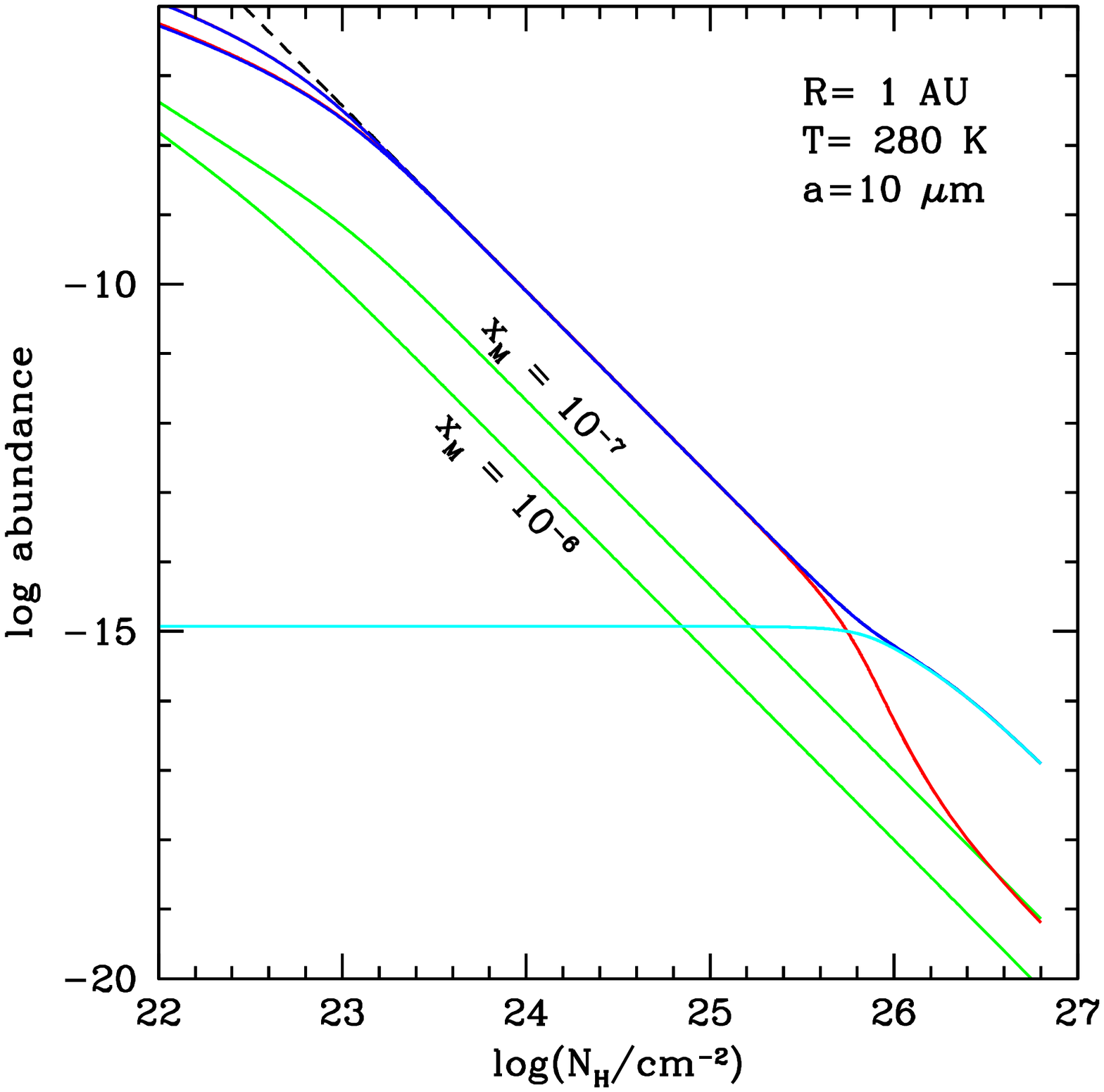} %{summ_1_xM_6_7.eps}
\end{minipage}
\caption{{\it Top panel:} Variation with  grain size, $a =  10, \, 100
\, \micron$. {\it Bottom panel:} Variation with heavy atom abundances, $x_M = 10^{-6},
\,10^{-7}.$ The curves have the same meaning as in Figure
\ref{fig:10m_1AU_10AU}.}
\label{fig:1AU_param}
\end{figure}

Figure~\ref{fig:1AU_param} displays the dependence on grain size
and heavy atom abundance obtained with the cubic Eq.~(\ref{eq:cubic})
at 1\,AU. The general increase of $x_e= x(M^+)$ in Figure~\ref{fig:1AU_param}
is in accord with the simple ionization theory of the previous
section, where $x_e \propto a$, using Eqs.~(\ref{eq:simplest}) and
(\ref{eq:grd}). Likewise, $x(g^-)$ decreases because, referring to
Eq.~(\ref{eq:grcharge}), it is inversely proportional to $a^2$. By
contrast, $x(m^+)$ is independent of grain size because according
to Eq.~(\ref{eq:molion}), the dependence on grain size of the two
factors, $k_d\,x_d$ and $x_e$, cancel. However, among all of the
species under discussion, $m^+$ is the only one sensitive to the
heavy element abundance, as shown in Figure~\ref{fig:1AU_param} and
predicted by Eq.~(\ref{eq:molion}).

\section{Application to Magnetic Disks}

We calculate the non-ideal MHD diffusivities in the low-ionization
limit given in Appendix B of Pinto, Galli \& Bacciotti~(2008), using
the rate coefficients in Pinto \& Galli~(2008a,b). The last equations
of  this appendix for the case where the charged particles are
electrons and heavy ions lead to these familiar forms for the Ohmic
(O), Hall (H) and Ambipolar Diffusion (AD) diffusivities,
\be
\eta_{\rm O} = \frac{1}{4 \pi r_e} \, \frac{\langle \sigma v \rangle_{e,n}}{x_e}
\label{eq:ohmic}
\ee
\be
\eta_{\rm H} = \frac{c B}{4 \pi e n_e}
\label{eq:hall}
\ee
\be
\eta_{\rm AD} = \frac{B^2}{4\pi\gamma_{i,n}\,\rho_i \,\rho_n} 
              \approx \frac{v_{\rm A}^2}{\nh \langle \sigma v \rangle_{i,n}\, x_i}
\label{eq:AD}
\ee
where $r_e = 2.818 \times 10^{-13}\,{\rm cm}$ is the classical
electron radius, $\langle \sigma v\rangle_{e,n}$ and $\langle \sigma
v \rangle_{i,n}$ are average rate coefficients for electron- and
ion-neutral momentum transfer, $c$ is the speed of light, $\rho_n
= \nh\,\mh$, $\gamma_{i,n} \rho_i\approx \nh \,x_i\,\langle \sigma
v \rangle_{e,n}$, and $v_{\rm A}$ is the Alfv\'en velocity,
\be
v_{\rm A} = \frac{B}{\sqrt{4 \pi \rho}} = 1.84 \times 10^{11}
\left(\frac{B}{G}\right) \left( \frac{\nh} {\rm cm^{-3}}\right)^{-1/2}\, \cmps
\label{eq:alfven}.
\ee
The temperature dependence of $\langle \sigma v \rangle_{e,\,{\rm
H}_2}$ is obtained from Table~1 of Pinto \& Galli (2008b); typical
values for this rate coefficient are $\sim 10^{-8} \, \ccps$. Now
using Eq.~(A.5) of Pinto \& Galli (2008b), $\langle
\sigma\,v\rangle_{M^+,\,{\rm H}_2} = 1.85 \times 10^{-9}\, \ccps$,
independent of temperature, taking the heavy atom mass to be
$25\,\mh$.

We evaluate the diffusivities and related quantities for the vertical
structure of a weakly-magnetized disk model around a T Tauri star
(Lizano et al.~2016). The disk is threaded by a poloidal magnetic
field dragged in from the parent core during the phase of gravitational
collapse (Shu et al.~2007).  This model has a a viscosity coefficient
$D=10^{-2.5}$ and a dimensionless mass-to-flux ratio $\lambda = {2
\pi G^{1/2} {\left(M_* +M_d\right)} / \Phi} = 12$, where the stellar
mass is $M_*=0.5~M_\odot$ and the disk mass is $M_d = 0.03~M_\odot$.
The disk radius is $R_d = 39.3$ AU and the mass accretion rate is
$\dot M = 10^{-8}~M_\odot~{\rm yr}^{-1}$.  The radial surface density
distribution is $\Sigma_R = 540.3\left( R/ {\rm AU}
\right)^{-3/4}$~g~cm$^{-2}$ (half above and half below the mid-plane).
The aspect ratio is $A = 0.033 \left(R/{\rm AU}\right)^{1/4}$. The
vertical magnetic field component is $B_z = 6.1 \left(R/{\rm
AU}\right)^{-11/8}\,{\rm G } $, and the radial component is $B_{R}
= 2 B_R^+ \left(\Sigma/ \Sigma_R \right)$ where  $B_R^+ = 1.742 \,B_z$ 
is the radial field at the disk surface. The disk is internally
heated by viscous and resistive dissipation and its surface is
irradiated by the central star.

\begin{figure*} %[!hbtp]
\includegraphics[scale=0.9]{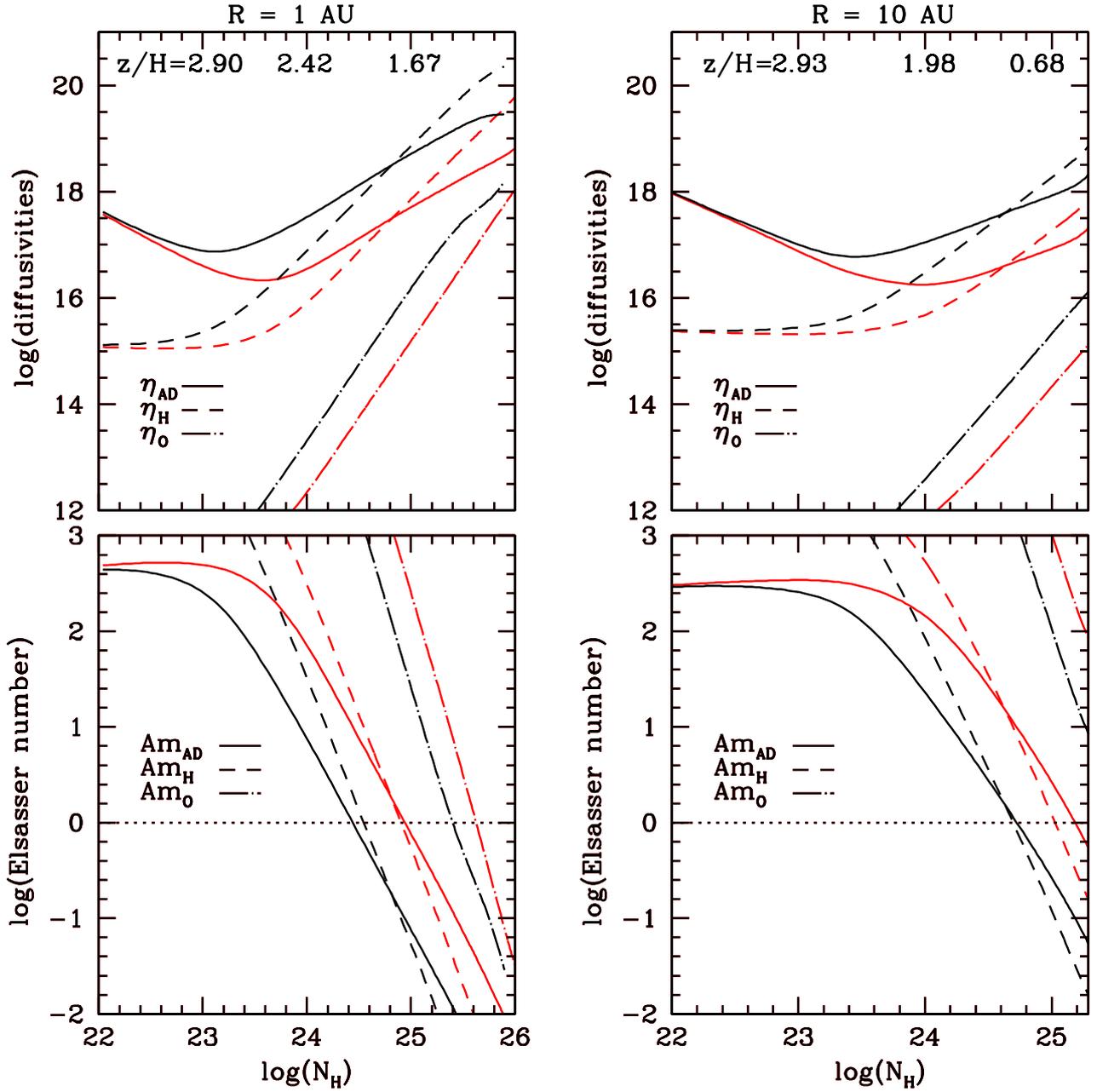} 
\caption{
Diffusivities and Elsasser numbers calculated for a weakly magnetized
T Tauri disk with a mass-to-flux ratio $\lambda = 12$, at disk radii $R
= 1$~AU ({\it left panels}\/) and $R = 10$~AU ({\it right panels}\/),
for grains with sizes $a = 10$~$\mu$m ({\it black curves}\/) and
$a=100$~$\mu$m ({\it red curves}\/).  The  solid lines correspond to
the Ambipolar diffusion terms ${\rm Am}_{\rm AD}$ and $\eta_{\rm
AD}$, the dashed lines correspond to the Hall terms ${\rm Am}_{\rm
H}$ and $\eta_{\rm H}$, and the dot-dashed lines correspond to the
Ohm terms ${\rm Am}_{\rm O}$ and $\eta_{\rm O}$. Note that the
mid-plane column density at 10 AU is smaller than the mid-plane
column density at 1~AU.  The top axis in the top panels correspond
to the number of scale heights is $z/z_{H}$ at each column density
(see text). For  these models, $H (1~{\rm AU}) = 2.25 \times 10^{-2}
\,$AU and $H (10~{\rm AU}) = 2.78 \times 10^{-1} \,$AU.}
\label{fig:Diff_Elsasser}
\end{figure*}

The focus is on intermediate altitudes below the FUV zone and
extending down close to the mid-plane. Here the dominant neutral
is molecular hydrogen, and the ionization parameter $\tilde{\zeta}$
varies approximately as $N_{\rm H}^{-8/3}$. The dependence of the
ionization fractions on column density is obtained with the theory
described in Sec.~2. In order  to better understand the numerical
values of the diffusivities, we also calculate the corresponding
Elsasser numbers using the Keplerian rotation frequency, $\Omega(R)
= 1.4 \times 10^{-7} \left({M_*}/{0.5 \, M_\odot}\right)^{1/2}
\left({R}/{\rm AU}\right)^{-3/2}~\ps$,
\be
{\rm Am}_{\rm O}= \frac{c_s^2}{\Omega \, \eta_{\rm O}} \hspace{0.35in}
{\rm Am}_{\rm H}= \frac{v_{\rm A}^2}{\Omega \, \eta_{\rm H}} \hspace{0.25in}
{\rm Am}_{\rm AD}= \frac{v_{\rm A}^2}{\Omega \, \eta_{\rm AD}},
\label{eq:elsasser}
\ee 
where $c_s$ is the isothermal sound speed for a molecular-hydrogen plus helium mixture,
$c_s = 6.00 \times 10^4\, (T/100\,{\rm K})^{1/2}$~cm~s$^{-1}$
%\be
%c_s = 6.00 \times 10^{4} \cmps \left(\frac{T}{100 \,K}\right)^{1/2},
%\ee
and the Alfv\'en velocity $v_{\rm A}$ is given above in
Eq.~(\ref{eq:alfven}). For the Hall term, we can also use a spatial
scale closely related to the Elsasser number, i.e., the so-called
Hall length (e.g., Lesur et al.~2014),
\be 
L_{\rm H} = \frac{\eta_{\rm H}}{v_{\rm A}} =\frac{v_{\rm A}}{\Omega}\,\frac{1}{{\rm Am}_{\rm H}}.
\label{eq:halllength}
\ee
Lesur et al.~(2014) interpret $L_{\rm H}$ as a measure of the spatial
range of the Hall effect. The unique characterizations of the
non-ideal MHD terms in Eqs.~(\ref{eq:elsasser}) and (\ref{eq:halllength})
arise from the different dependences of the original diffusivities,
Eqs.~(\ref{eq:ohmic}),\, (\ref{eq:hall}) and (\ref{eq:AD}), on
ionization fraction and magnetic field.

Figure~\ref{fig:Diff_Elsasser} plots the three diffusivities and
Elsasser numbers vs.~vertical column density for two radii, 1\,AU
and 10\,AU, and two grain sizes, $a = 10$ (black lines) and $100
\, \micron$ (red lines). Focusing on the Elsasser numbers on the
lower panels of Figure~\ref{fig:Diff_Elsasser}, they all decrease
monotonically with $N_{\rm H}$. At high altitudes, their numerical
values satisfy the sequence: ${\rm Am}_{\rm AD} < {\rm Am}_{\rm H}
< {\rm Am}_{\rm O}$, but this can change deeper down. The usual
interpretation of the two Elsasser numbers, ${\rm Am}_{\rm O}$ and
${\rm Am}_{\rm AD}$, comes from shearing-box simulations with large
plasma $\beta$ (weak fields). It is based on the idea that large
${\rm Am}_{\rm O}$ guarantees that the ionized disk plasma is well
coupled to the magnetic field, whereas large ${\rm Am}_{\rm AD}$
guarantees that the ions are well coupled to the neutrals. Many
authors found that the MRI is active if the Elsasser numbers are
$>100$, but others have advocated the condition $>1$ (e.g., Bai \&
Stone 2011, Bai 2015). If we were to adopt the value of 1, then
Figure~\ref{fig:Diff_Elsasser} suggests that ${\rm Am}_{\rm O}$ is
probably  large enough to provide good field-plasma coupling over
much of the disk between 1 and 10\,AU, but ${\rm Am}_{\rm AD}$ is
not large enough to provide adequate coupling of ions and neutrals
over much of the disk. The steep slopes of the Elsasser numbers in
this figure suggest that such deductions may depend on the choice
of the critical number for the MRI.  However, there will still be
dead zones in the present case. Furthermore, the size of the dead
zone is smaller for larger grains.  Flock et al.~(2012) carried out
a global zero net-flux simulation that indicates that the MRI can
be sustained with Elsasser numbers as low as 0.1. Even with this
small value, however, dead zones still arise at 1\,AU in the present
model. In this context, it should be noted that the weakly-magnetized
T Tauri disk model star considered here is fairly thick with a
mid-plane thickness of $540\, {\rm g \, cm^{-2}}$; thinner disks
may well have no dead zones.  The upper axis of the diffusivity
panels in the figure show the number of scale heights at a given
column density. As in the case of  isothermal disks, we chose the
scale height $H$ as the value of the height $z$ that contains 68\%
of the surface density, measured from the mid-plane. For $R=1$ AU,
$H= 2.25 \times 10^{-2}$ AU, such that $z/H = 3.55, 2.90, 2.42$ and
1.67 for $\log N_H = 22, 23, 24$ and $25 \, {\rm cm}^{-2}$. For
$R=10$ AU, $H= 2.78 \times 10^{-1}$ AU, such that $z/H = 4.00, 2.93,
1.98$ and 0.68 for $\log N_H = 22, 23, 24$ and $25 \, {\rm cm}^{-2}$.

The present model considers a single ionization source, stellar
X-rays, for the region below the FUV zone. As noted in Secs. 1 \&
2, the X-ray ionization rate becomes uncertain beyond a certain
depth, e.g., $\log N_{\rm H} = 25.25$ at 1\,AU. There has been much
speculation on the possibility of other ionization sources close
to the mid-plane. In addition to stellar and galactic cosmic rays,
locally-generated energetic particles may generate sufficient
ionization to reduce the amount of MRI-dead material (e.g., Turner
\& Drake~2009; see also Gounelle~2015 for the role of radioactive
nuclides and Ilgner \& Nelson~2008 for the potential role of mixing).
MRI turbulence itself may play a role here, as discussed recently
by Inutsuka \& Sano~(2005) and Okuzumi \& Inutsuka~(2015).

\begin{figure} %[!hbtp]
\centering
\includegraphics[angle=0,width=\columnwidth]{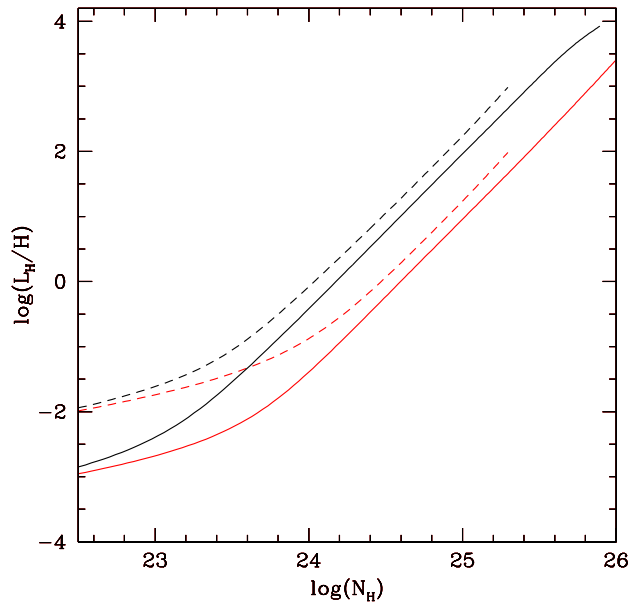} \caption{Ratio
of the Hall length to the disk scale height $H$ as function of
vertical column density at $R=1$~AU ({\it solid curves}\/) and
$10$~AU ({\it dashed curves}\/), for $a=10$~$\mu$m ({\it black
curves}\/) and $100$~$\mu$m ({\it red curves}\/).}
\label{fig:Hall_length}
\end{figure}

The Hall length $L_{\rm H}$ as function of vertical column density
is shown in Figure~\ref{fig:Hall_length} for $R=1$ and 10~AU and
$a=10$ and 100~$\mu$m.  Eq.~(\ref{eq:halllength}) for the Hall
length indicates that it is inversely proportional to ${\rm Am}_{\rm
H}$. The small values of ${\rm Am}_{\rm H}$ approaching the mid-plane
in Figure~\ref{fig:Diff_Elsasser} mean that $L_{\rm H}$ is much larger
than the disk scale-height at the disk mid-plane.  This is consistent
with the generally accepted idea that the PPDs are significantly
affected by the non-ideal MHD Hall term. The Hall length in these
models is larger than the values obtained by Lesur et al.~(2014)
in their Figure~3, because of the different disk models, and also
because of their higher ionization fraction due to the inclusion
of cosmic ray ionization.

{The discussion of Figure \ref{fig:Diff_Elsasser} emphasizes the occurrence and size of MRI dead zones. 
A corollary to the conclusion that thicker disks can have dead zones deep down, is that the MRI may still be active 
over a large volume of a T Tauri star disk. The turbulence associated with the MRI has long been considered to 
play a key role in angular momentum transport (e.g., Hawley \& Stone 1998, Bai \& Goodman 2009 and Mohanty et al. 2013), 
where the ionization is critical for the MRI. More recently, the transport of angular momentum has focused on the important 
role of disk winds (e.g.,  Bai et al. 2016, Bai  2016, Bai \& Stone 2017).  Although the thermal-chemical properties 
of the upper part of the disk (including the ionization) are important in this connection,  the present theory of of deep-down ionization 
will not be relevant unless the base of the wind extends well below the FUV layer.   }

\section{Summary}

By applying the methodology of Oppenheimer \& Dalgarno~(1974) (see
also Ilgner \& Nelson~2006a,b and Bai \& Goodman~2009), we have
obtained a simple description of the deep-down X-ray ionization of
protoplanetary disks. The key step is to treat a small set of generic
species, molecular ions like $\hcop$ and $\hthreeop$ (represented
by $m^+$), heavy atomic ions like Na$^+$ and Mg$^+$ (represented
by $M^+$), and charged grains ($g^-$), as well as electrons ($x_e$).
Below the FUV irradiated surface region, the ionization is described
by a cubic equation in $x_e$. Over much of the inner disk, the
vertical variation of the ionization follows simple closed-form
expressions that are essentially power laws in vertical column
density. We have illustrated the theory by calculating the Elsasser
numbers for the standard non-ideal MHD effects for the weakly-magnetized
T Tauri star disk model of Lizano et al.~(2016). We foresee other
applications of the theory to 2-d and 3-d MHD modeling of magnetic
disks.

Another key aspect of the theory is the inclusion of the reactions
whereby atomic ions transfer their charge to grains. This process
has been ignored in some MRI simulations, with the result that $x_e
= (\zeta_X/\beta n_{\rm H})^{1/2}$, where $\beta$ is the dissociative
recombination rate coefficient for molecular ions. This expression
produces too small values of $x_e$ near the mid-plane of moderately
thick disks (see instead eq.~\ref{eq:simplest}).

The present application would seem to suggest that fairly thick
disks, approaching the thickness of the MMSN have dead zones even
with X-ray ionization. This raises an important limitation of the
theory in that the level of ionization deep down, e.g., beyond log
$N_{\rm H} = 25$, is uncertain. This is mainly due to the limitations
in the Monte Carlo propagation of the scattered X-rays and our
incomplete understanding of the blocking of galactic cosmic rays
by the angular-dependent stellar wind. And beyond these major
uncertainties, there may be additional internal sources of ionization,
such as the MRI itself.

\section*{Acknowledgments}
S. L . acknowledges support from PAPIIT-UNAM IN105815 and CONACytT 238631.

\label{lastpage}
\end{document}